\documentclass[aps,prappl,reprint,superscriptaddress]{revtex4-2}
\DeclareUnicodeCharacter{03BC}{\ensuremath{\mu}}
\usepackage{graphicx}
\usepackage{lipsum}
\usepackage{amsmath}
\usepackage{siunitx}

\begin{document}

%\preprint{}

\title{Native-oxide-passivated trilayer junctions for superconducting qubits}


\author{Pankaj Sethi}
\email{pankaj.sethi@vtt.fi}
\author{Om Prakash}
\altaffiliation{Present address: Department of Physics and Astrophysics, University of Delhi, New Delhi 110007, India}
\author{Jukka-Pekka Kaikkonen}
\author{Mikael Kervinen}
\author{Elsa T. Mannila}
\author{Mário Ribeiro}
\altaffiliation{Present address: Arctic Instruments, Tekniikantie 14, FI-02150 Espoo, Finland}
\author{Debopam Datta}
\author{Christopher W. Förbom}
\author{Jorden Senior}
\author{Renan P. Loreto}
\author{Joel Hätinen}
\author{Klaara Viisanen}
\affiliation{VTT Technical Research Centre of Finland Ltd, QTF Centre of Excellence, P.O. Box 1000, FI-02044 VTT, Finland}
\author{Jukka I. Väyrynen}
\affiliation{Department of Physics and Astronomy, Purdue University, West Lafayette, Indiana 47907, USA}
\author{Alberto Ronzani}
\author{Antti Kemppinen}
\author{Visa Vesterinen}
\altaffiliation{Present address: Arctic Instruments, Tekniikantie 14, FI-02150 Espoo, Finland}
\author{Mika Prunnila}
\author{Joonas Govenius}
\altaffiliation{Present address: Arctic Instruments, Tekniikantie 14, FI-02150 Espoo, Finland}
\email{joonas@arcticinst.io}
\affiliation{VTT Technical Research Centre of Finland Ltd, QTF Centre of Excellence, P.O. Box 1000, FI-02044 VTT, Finland}


\date{\today}

\begin{abstract}
Superconducting qubits in today's quantum processing units are typically fabricated with angle-evaporated aluminum--aluminum-oxide--aluminum Josephson junctions. However, there is a need for  higher yield and tighter parameter control when scaling up the number of qubits and junctions into tens of thousands and beyond. Fabrication methods based on subtractive patterning of superconductor--insulator--superconductor trilayers, used for more classical large-scale Josephson junction circuits, could provide the solution but they in turn often suffer from lossy dielectrics incompatible with high qubit coherence. 
In this work, we utilize native aluminum oxide as a sidewall passivation layer for junctions based on aluminum--aluminum-oxide--niobium trilayers, and use such junctions in qubits. We design the fabrication process such that the few-nanometer-thin native oxide is not exposed to oxide removal steps that could increase its defect density or hinder its ability to prevent shorting between the leads of the junction. With these junctions, we design and fabricate transmon-like qubits and measure time-averaged coherence times up to \SI{30}{\micro\second} at a qubit frequency of \SI{5}{\giga\hertz}, corresponding to a qubit quality factor of one million. Our process uses subtractive patterning and optical lithography on wafer scale, enabling high throughput in patterning. This approach provides a scalable path toward fabrication of superconducting qubits on industry-standard platforms.
\end{abstract}

%\keywords{}

\maketitle

\section{Introduction}

Superconducting tunnel junctions are a key element providing non-linearity to many circuits in the field of low-temperature sensors and electronics. Several approaches to their fabrication exist, optimized for different applications. Most of the approaches fall into the category of angle-evaporation combined with lift-off \cite{dolan_offset_1977, potts_novel_2001} or superconductor--insulator--superconductor trilayers combined with subtractive patterning \cite{gronberg_fabrication_2007, tolpygo_advanced_2019, dorojevets_16-bit_2013}. Superconducting qubits use the former almost exclusively because, so far, approaches based on lift-off and electron-beam lithography have produced more ideal low-loss junctions in the sense of introducing less material capable of absorbing energy at the qubit frequency. Furthermore, so far, high-gate-fidelity superconducting quantum processing units have contained at most a few hundred junctions, such that parameter control achieved with angle-evaporation and lift-off has been acceptable. To our knowledge, the best reported wafer-level failure rates are in the range of 0.1 to 1\%, when defining only short and open circuits as failures \cite{kreikebaum2020improving, muthusubramanian2024wafer}. The trilayer-based approach is, on the other hand, prevalent in more classical applications of superconducting junctions, such as magnetometers, current amplifiers, and single-flux quantum circuits. These applications often require a very large number of junctions and tight parameter tolerances \cite{dorojevets_16-bit_2013}, and therefore the superior yield, reproducibility, and lack of aging in the approach based on trilayers and subtractive patterning are essential. Superconducting quantum processors are expected to reach a similar scale in terms of number of junctions in the next years, therefore making it a pressing problem to develop trilayer-based junctions with qubit-compatible level of microwave loss \cite{NakamuraAPL2011, weides_phase_2011, kim2021enhanced, wan_fabrication_2021, anferov_improved_2024}. There are also other ways of eliminating lift-off from the conventional qubit junction process, such as the two-step angle-evaporation-free approach \cite{wu_overlap_2017}, which can also be adapted to subtractive patterning compatible with large-scale fabrication \cite{stehli_coherent_2020, van_damme_advanced_2024}.

The conventional pillar-junction approach to trilayer-based junctions \cite{gronberg_fabrication_2007, tolpygo_advanced_2019} results in a large volume of passivation material in the vicinity of the junction, and the materials conventionally used for the passivation are anodized niobium pentoxide \cite{dolata_sub_1999} and deposited amorphous insulators with poor microwave loss tangent $\tan \delta$. For example, typical silicon dioxide, SiO$_2$, deposited by plasma-enhanced chemical vapor deposition has a $\tan \delta$ no better than $10^{-3}$ at few-gigahertz frequencies, especially when deposited at low enough temperatures to not damage aluminum oxide tunnel barriers \cite{li_improvements_2013}. This implies that no reasonable superconducting qubits can be fabricated with the conventional pillar-junction approach. A more recent approach, the side-wall spacer passivated sub-\SI{}{\micro\meter} Josephson junction fabrication process (SWAPS) differs from the pillar-junction approach in that the passivation layer is etched away almost completely and only remains on the sidewalls of the patterned trilayer \cite{gronberg_side-wall_2017}. 
Similar junctions with anodized niobium pentoxide on the sidewalls can also be created with anodization \cite{dolata_sub_1999}.
Nevertheless, because the remaining sliver of passivation on the sidewalls is located at a point of high electric field and is made of an amorphous insulator with poor $\tan \delta$, the microwave loss introduced by the spacer in the SWAPS and anodized solutions is still too large to build state-of-the-art superconducting qubits. The removal of spacer oxide from SWAPS junctions has been demonstrated \cite{anferov_improved_2024}. However, in our experience, it is difficult to remove the spacer oxide reliably without causing undesirable degradation of the aluminum-based tunnel barrier or the surrounding niobium parts. These potential side-effects tend to lead to poor critical-current control and excess microwave loss, in addition to a suspended structure which may ultimately limit post-processability and yield. Other trilayer-based approaches resulting in suspended structures include junctions made from fully epitaxial NbN/AlN/NbN trilayers, including demonstration of $T_1 \approx 0.3$--$0.5\,\mu\text{s}$ already in Ref.~\cite{NakamuraAPL2011} and later $16\,\mu\text{s}$ in Ref.~\cite{kim2021enhanced}. 

Therefore, for qubit-compatible trilayer-type junctions, the key challenge is to passivate the bottom metal of the trilayer such that a top contact can be reliably formed, while simultaneously minimizing the volume of material with poor loss tangent in areas of high electric field. More quantitatively, in order for the passivation to not limit the overall qubit performance, the product $p_\mathrm{pass.} \tan \delta$ should be small in comparison to the inverse of the effective qubit quality factor $Q = 2 \pi f_q T_1$, where $p_\mathrm{pass.}$ is the participation ratio of the passivation, $f_q$ is qubit frequency, and $T_1$ is the energy relaxation time of the qubit. For state-of-the-art superconducting qubits, $1/Q$ is of order $10^{-7}$ \cite{deng_titanium_2023, tuokkola_methods_2025}. For the SWAPS process and typical transmon qubit parameters, $p_\mathrm{pass.}$ is of order $10^{-4}$ or greater and $\tan \delta$ is $10^{-2}$ or greater, and therefore $p_\mathrm{pass.} \tan \delta$ is limited to no better than $10^{-6}$.

In this work, we introduce a native-oxide-passivated trilayer junction (NAPA junction) that can be fabricated reliably and in which the contribution of passivation to $1/Q$ is negligible under reasonable assumptions. Specifically, we use a SWAPS-like structure based on an Al-Al$\mathrm{O}_x$-Nb trilayer with native aluminum oxide as the sidewall passivation, which remains intact thanks to the introduction of an additional via into the structure. The contribution of the passivation layer to $p_\mathrm{pass.} \tan \delta$ is small, as long as we assume that the loss-tangent of the native-oxide passivation is not significantly worse than that of the tunnel barrier. Recently, an approach with a native aluminum oxide passivation layer has been used to fabricate junctions patterned into an Al-Al$\mathrm{O}_x$-Al trilayer~\cite{mechold_towards_2024} that also use aluminum native oxide as passivation. However, NAPA junctions differ in crucial details of how electrical contact is made between the counter electrode of the trilayer and the top wiring layer. In particular, thanks to an additional via described in more detail below, the NAPA junction fabrication process avoids exposing the native-oxide sidewall passivation to ion milling or other aggressive oxide removal steps during the fabrication process.

Using optically-patterned NAPA junctions with a lateral size of approximately 600 nm, we fabricate superconducting transmon qubits and demonstrate time-averaged $T_1$ time up to \SI{30}{\micro\second}, and average qubit quality factors of 0.5 million over several chips and wafers, with the best devices achieving quality factors close to one million.  We also discuss the implications of these results to the question of the role of subgap leakage in qubits with Nb-based tunnel junctions.

\section{Junction structure and fabrication}

\begin{figure*}[tb]
\includegraphics{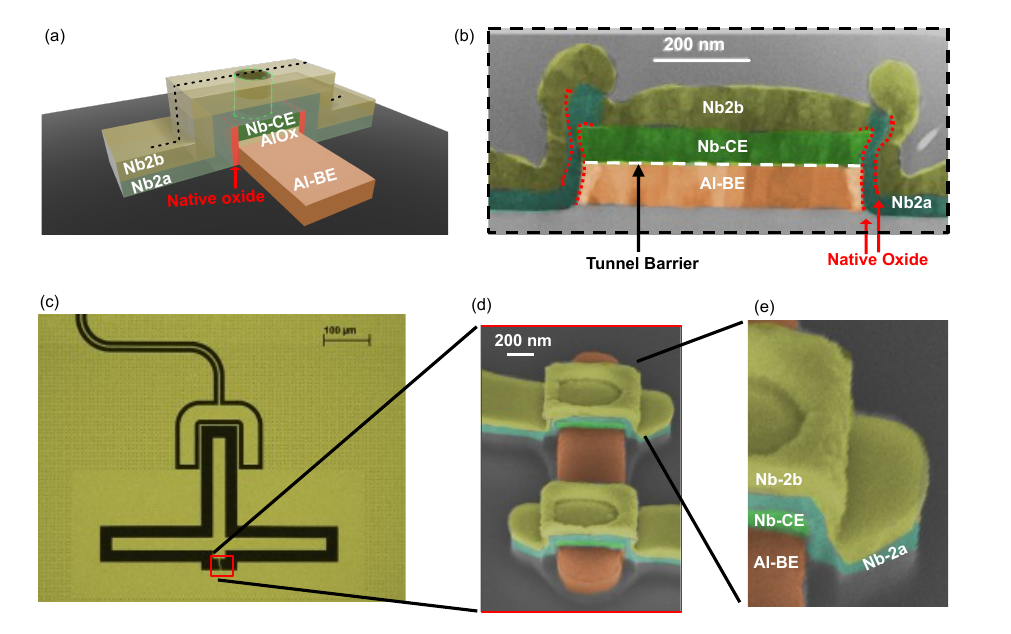}%
\caption{\label{fig:schematic}
(a) Schematic illustration of a NAPA junction and (b) false-colored cross-sectional transmission electron micrograph of a typical fabricated NAPA junction along the dashed black line shown in panel (a). The aluminum base electrode (Al-BE, orange), tunnel barrier (white dashed line) and niobium counter electrode (Nb-CE, light green) are insulated from the Nb wiring layer (Nb2a and Nb2b) by a native oxide passivation layer (red dashed line). The via providing galvanic contact between Nb2 and Nb-CE is on top of the junction.(c) False-color scanning electron micrograph of fabricated Xmon-type shunt capacitor patterned into Nb2, capacitively coupled to a readout resonator (partially visible at the top). (d) Close-up of two NAPA junctions in series, with a small island of Al-BE in between. (e) Further zoomed in close-up of the junction in panel (d).
}
\end{figure*}

Figure~\ref{fig:schematic}(a) and (b) show the conceptual structure and a cross-sectional transmission electron micrograph of a typical NAPA junction.  Let us first describe the final structure, before describing how it is fabricated: The tunnel barrier of the junction is defined by a trilayer of aluminum base electrode (Al-BE), aluminum oxide tunnel barrier and a niobium counter electrode (Nb-CE). The trilayer lies directly on a high-resistivity silicon substrate, without silicon oxide or other lossy dielectric layers in between. The sidewalls of the patterned trilayer are covered by a native-oxide spacer, which is a relatively low-loss dielectric and electrically insulates Al-BE from the Nb wiring layer (Nb2). The aluminum bottom electrode Al-BE acts as one of the leads for connecting the junction to other circuit elements. The wiring layer Nb2 acts as the second lead and consists of two sublayers (Nb2a, Nb2b), with a via etched through Nb2a and partially into the counter electrode (Nb-CE), in order to provide electrical contact from Nb2 to Nb-CE. The sublayers Nb2a and Nb2b are galvanically connected to each other almost everywhere. Along the vertical sidewalls of the junction a layer of native niobium oxide separates Nb2a and Nb2b [Fig.~\ref{fig:schematic}(b), Fig.~\ref{fig:junction}(c)] but that separation serves no functional purpose.

The advantage of this additional complexity in forming Nb2 is that it is possible to avoid exposing the thin native-oxide passivation of the sidewalls to oxide removal steps, such as ion milling. More specifically, the main steps of the fabrication process are:
First, we deposit the unpatterned trilayer on a high-resistivity silicon substrate, using a single tool and without breaking vacuum between deposition of the constituent layers, except for in-situ oxidization of the tunnel barrier at a controlled oxygen pressure.
Second, we pattern the trilayer using optical lithography and reactive ion etching, after which the native-oxide passivation grows on the patterned Al-BE and Nb-CE, upon exposure to oxygen when the wafer is removed from the loadlock of the etching tool. We make no effort to control the wafer temperature or the the introduction of oxygen and humidity at that point, which may impact the quality of the native oxide that forms. There are no major hurdles to controlling these aspects of the oxidization process in the future.
Third, we deposit Nb2a, without ion milling or other oxide removal steps. Therefore, Nb2a is not at this point electrically connected to the trilayer counter electrode Nb-CE.
Fourth, we optically pattern and reactively etch a via through Nb2a and partially into the counter electrode (Nb-CE). Once the wafer is removed from the etching tool, a native oxide forms on the part of Nb-CE exposed through the via, but is removed in the next step.
Fifth, we deposit Nb2b, this time with in-situ ion milling preceding the deposition to remove oxides from all exposed surfaces. Thus, inside the via, we form a galvanic connection between Nb2b and Nb-CE, and also connect Nb2b to Nb2a almost everywhere, without native niobium oxide in between.
Finally, we optically pattern and etch both Nb2 sublayers, and the remaining Nb-CE outside the junction area, the same way as in the conventional SWAPS process \cite{gronberg_fabrication_2007}. See Appendix~\ref{appendix:fabrication} for more details on fabrication.

The introduction of a via and splitting of the deposition of Nb2 into two sublayers allow avoiding oxide removal steps when the native-oxide passivated sidewalls of the trilayer are exposed. Instead, surface oxide removal using ion milling is delayed until the sidewalls of the trilayer are protected by Nb2a. Similarly, depositing Nb2a first protects the silicon substrate surface from ion milling. This is a key difference to the method described in Ref. \cite{mechold_towards_2024}. In fact, no surfaces are ion milled in our approach, except Nb surfaces that are later covered by another Nb layer. All other surfaces are exposed to only reactive ion etching, which operates at lower ion energies than physical ion milling due to its chemical nature. Consistent with this, quarter wave resonators on the same wafers as the qubits reach internal quality factors above ca. half a million, as discussed later.  

Figure~\ref{fig:schematic}(c) shows a false-colored scanning electron micrograph of NAPA junctions used in a transmon-like qubit, with a large shunt capacitor in the Xmon geometry \cite{barends_coherent_2013}. Panels (d) and (e) show detailed close-ups of the junctions. The qubit is capacitively coupled by the claw-shaped coplanar capacitor to a readout resonator (partially visible towards the top). Here, we have used two NAPA junctions in series, in contrast to conventional transmons. The double-junction design allows keeping the material left in Al-BE small. It also enables the use of the same material (the Nb2 wiring layer for the device in Fig.~\ref{fig:schematic}) for both the X-shaped island electrode and the ground electrode, without requiring large parasitic junctions to connect the qubit junction to the ground plane. The circuit model for the double-junction qubit is discussed in the next section.

We utilize the native aluminum oxide, that forms on the etched Al-BE sidewalls to passivate the bottom electrode and prevent leakage currents between the bottom electrode and top electrode. An unavoidable native niobium oxide layer also grows on the Nb counter-electrode but is surrounded on both sides by galvanically-connected top electrode metal (Nb-CE and Nb2a). By using native aluminum oxide as passivation, we expect to construct higher-coherence-time superconducting qubits than with other trilayer-type junctions, such as SWAPS or pillar-type junctions. In more exact terms, $p \tan \delta$ can be reduced, despite the tendency of $p$ to increase with decreasing thickness of the sidewall passivation. In the case of aluminum oxide, $\tan \delta$ of approximately $2 \times 10^{-6}$ has been reported \cite{sliwa_characterization_2023, murray_material_2021, mamin_merged-element_2021}.  

We also argue that the native oxide of any material suitable for forming high-quality tunnel barriers is likely to work well as sidewall passivation, assuming that the native oxide formation is controlled such that its microwave loss tangent is similar to that of the tunnel barrier. Microwave losses in the tunnel barrier will then dominate because the participation ratio of the sidewall passivation is always smaller than that of the tunnel barrier, if lateral dimensions are larger than film thicknesses.
Quantitatively, we estimate a ratio of 4:1 between the tunnel barrier capacitance and the sidewall passivation capacitance, for our specific geometry. Given the thin dielectric thicknesses and large aspect ratios, we can estimate the capacitances using parallel-plate capacitor approximations: $\varepsilon_{\mathrm{AlO}_x} \times (600\ \mathrm{nm})^2 / 1.5\ \mathrm{nm}$ for the tunnel barrier and $\varepsilon_{\mathrm{AlO}_x} \times 2(600\ \mathrm{nm})(100\ \mathrm{nm}) / 2\ \mathrm{nm}$ for the sidewall passivation, with $\varepsilon_{\mathrm{AlO}_x}$ the dielectic constant of aluminum oxide. In the future, the lateral dimensions are likely to be scaled down but so is the thickness of Al-CE, so the overall ratio of these capacitances is not likely to change dramatically.

\section{Junction characterization}

\begin{figure}[tb]
\includegraphics{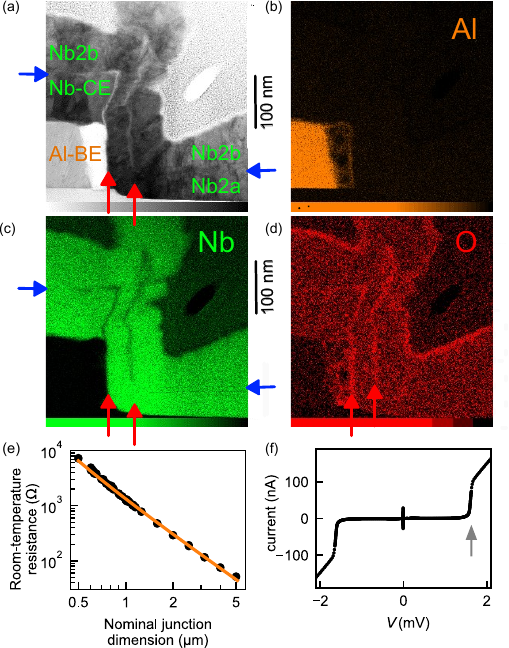}%
\caption{\label{fig:junction}(a)-(d) TEM image and EDS analysis spectra on the junction cross-section shown in Fig.~\ref{fig:schematic}, with the relative concentrations (b) aluminum, (c) niobium, and (d) oxygen. Native oxide layers, indicated by red arrows, are visible as areas with light contrast in panels (a) and (d) and dark contrast in panel (b). Note the absence of oxide between Nb-CE and Nb2b on the left and between Nb2a and Nb2b on the right, indicated by blue arrows, implying a galvanic contact. (e) The resistances of test junctions versus nominal junction size $d$, along with a fit  showing the expected scaling with junction area (solid orange line). (f) Low-temperature current-voltage characteristics of a test junction with $d=$ 629 nm measured at $40~$mK, showing a supercurrent branch at zero voltage and a sharp current rise at a voltage 1.6 mV indicated by arrows, corresponding to the sum of the superconducting gaps of Nb and Al, confirming the Josephson behaviour of NAPA junctions.}
\end{figure}

To characterize the fabricated junctions in more detail, we perform energy dispersive spectroscopy analysis (EDS) on the junction cross-section shown in Fig.~\ref{fig:schematic}. Fig.~\ref{fig:junction}(a) shows a TEM image along with EDS maps of the Nb, Al and O content. The native oxide layers are visible as layers with bright contrast in the O map and dark contrast in the Nb map. The native Al oxide is visible between Al-BE and Nb2a. In contrast, native oxides are removed between Nb-CE and Nb2b on top of the junction, and between Nb2a and Nb2b in the part of wiring layer laying directly on the substrate. In addition, vertical native Nb oxide is visible between Nb2a and Nb2b in both the O and Nb maps, as expected due to directionality of the argon milling before deposition of Nb2b. The tunnel barrier between Al-BE and Nb-CE is less than 2 nm thick and not clearly visible in these images.

We also find that the microstructure of Al-BE near the sidewalls differs significantly from the bulk, as indicated by the bright contrast at the edge of Al-BE in Fig.~\ref{fig:junction}(a), and by the mix of aluminum and oxygen in panels (b) and (d). This change of microstructure is likely caused by the reactive ion etching step used to pattern the trilayer, or corrosion due to residual chlorine or fluorine left over from the etching process. Due to these changes in Al-BE sidewall microstructure, we cannot quantitatively estimate the thickness of the passivation layer, and thus we cannot meaningfully estimate the loss tangent of the sidewall passivation alone, even though qubit lifetimes allow us to give an upper bound for the losses caused by the junction as a whole. We also note that the oxygen concentration in the Nb layers is visibly higher than the oxygen content in Al. 

We fabricate square test junctions with nominal lateral dimension $d$ between 500 nm and \SI{5}{\micro\meter} and measure their resistance at room temperature with an automated wafer prober in a four-probe configuration. Details of wafers are provided in Appendix~\ref{appendix:devices-wafers}. The measured resistance $R$, shown for a test chip on wafer A in Fig.~\ref{fig:junction}(e), scales as expected with the junction area as $R = RA / (d-l)^2$, where $RA = \SI{1100}{\ohm\micro\meter^2}$ is the mean resistance-area product across the wafer and $l = \SI{90}{\nano\meter}$ describes the reduction of the effective junction dimension from the nominal value on the photomasks. Direct proportionality of the junction conductance with the effective junction area confirms that the measured conductance arises from the tunnel barrier itself, and not from leakage through the sidewalls. For nominally identical junctions with a design diameter of 600 nm on wafer A, SEM measurements on 50 junctions distributed over the wafer give $d=600 \pm 20~$nm (1$\sigma$). The corresponding normal state $RA = 1100 \pm 100$ $\Omega\,\mu\text{m}^{2}$. The room temperature junction resistance varies by $\pm5\%$ ($\pm10\%$ full range) across the wafer, consistent with the observed dimensional spread due to lithography.

In Fig.~\ref{fig:junction}(f), we show low-temperature current-voltage (IV) characteristics of a single test junction with $d=629~$nm and normal-state resistance of \SI{10}{\kilo\ohm} from wafer B, measured in a dilution refrigerator at 40 mK. We observe a supercurrent branch at zero voltage with a switching current of 30 nA, and a sharp rise in current at a voltage bias of $(\Delta_{\mathrm{Nb}} + \Delta_{\mathrm{Al}})/e = $1.6~mV, where $e$ is the electron charge and $\Delta_{\mathrm{Nb}}$ ($\Delta_{\mathrm{Al}}$) is the superconducting gap in the niobium (aluminum) layer. These findings confirm that the device operates as a Josephson junction between Al and Nb. We discuss the subgap IV characteristics and their relation to qubit coherence below and in Appendix~\ref{appendix:dc}.

\section{Qubits with NAPA junctions}

In this work, we mainly study transmon-like qubits with two Josephson junctions in series, which allows us to avoid large parasitic junctions without introducing additional processing steps. For comparison, we also fabricate conventional fixed-frequency transmons, where large parasitic junctions are required to connect the small-critical-current transmon junction to the ground plane, since we wish to pattern both electrodes of the qubit shunt capacitor into the same metal layer.

Due to our relatively large 10 fF junction capacitance, the two-junction qubits have extremely low charge sensitivity of only a few hertz (see Appendix~\ref{appendix:double-jj}), despite having a small ($\lesssim$2 $\mu$m) superconducting island between the junctions. The lowest-energy excitations of this qubit type are transmon-like, with energy oscillating between the main shunt capacitor $C_\mathrm{S}$ and the effective inductor formed by the two junctions in series. We estimate $C_\mathrm{S} \approx 100\ \mathrm{fF}$ using electrostatic simulations. In the hypothetical case of small junction capacitances $\sim$1~fF, typical for Al--Al$\mathrm{O}_x$--Al junctions with lateral dimensions on the order of 100 nm, the charging energy of the island would be high and the two junctions would form a single Cooper pair transistor \cite{schreier_suppressing_2008}, making the effective inductance highly dependent on offset charge. 
Even in that case, however, coplanar shunt capacitance could be added to restore excellent charge insensitivity. 

We also fabricate devices where the roles of Al-BE and Nb2 in the design are inverted ("Al qubits"): That is, the qubit shunt capacitor, ground plane and resonators are patterened into Al-BE, while the small island between two junctions is Nb. We emphasize that this is a layout difference in the lithographic patterns only, and does not imply any changes in the processing steps. The individual junctions themselves and their native aluminum oxide passivation in particular remain nominally identical for the "Al qubits" and the "Nb qubits", except for uncorrelated minor fabrication differences between wafers (Appendix~\ref{appendix:devices-wafers}). Regardless of qubit type, each qubit is capacitively coupled to a $\lambda/4$ coplanar waveguide readout resonator with resonant frequency between 6 GHz and 7 GHz. Each chip contains three to seven individual qubits, with readout resonators inductively coupled to a common feedline for frequency-multiplexed readout and qubit drive. Appendix~\ref{appendix:devices-wafers} gives more details on the characterized devices and wafers.

\begin{figure}[tb]
\includegraphics{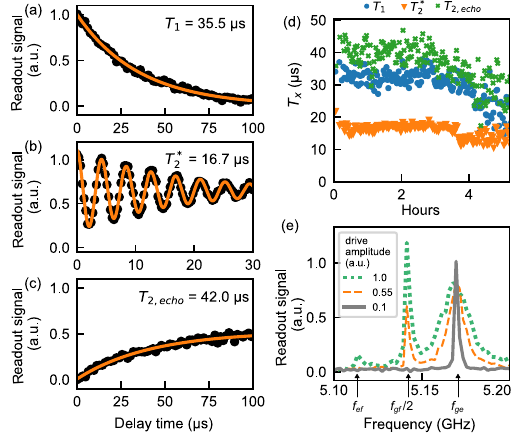}%
\caption{\label{fig:single-qubit}(a) Decay of a qubit at frequency $f_{ge} = 5.17$~GHz from the excited state, measured data (points) fit to exponential decay with $T_1=\SI{36}{\micro\second}$ (solid line), corresponding to a qubit quality factor $Q = 2 \pi f T_1 \approx $ 1 million. (b) Ramsey and (c) Hahn echo experiments fit to a decaying sinusoid and exponential decay, respectively. (d) $T_1$, $T_2^*$, $T_{2,\mathrm{echo}}$ fluctuate over time when monitored for several hours. (e) Two-tone spectroscopy experiment, with qubit transitions at frequencies $f_{ge}$, $f_{ef}$ and $f_{gf}/2$ indicated by arrows. We extract an anharmonicity of \SI{-56}{\mega\hertz}.}
\end{figure}

We characterize qubits using standard pulse sequences in a dilution refrigerator setup at 10 mK (see Appendix~\ref{appendix:measurement-setup} for details). In Fig.~\ref{fig:single-qubit}, we show one of the best lifetime measurements, for a double-junction Nb qubit with nominal junction size $d=(600\ \mathrm{nm})^2$ and qubit frequency equal to 5.17 GHz. The decay as a function of delay between excitation and readout is exponential with a time constant $T_1 = \SI{36}{\micro\second}$. Figures \ref{fig:single-qubit}(b) and (c) show results of Ramsey and Hahn echo experiments, respectively, which are well described by exponentially decaying sinusoids with time constants $T_2^* = \SI{17}{\micro\second}$ and $T_{2,\mathrm{echo}} = \SI{42}{\micro\second}$, respectively.  The y-axis, ranging from 0 to 1, approximately represents the excited-state population of the qubit but is not precisely calibrated.
When monitored repeatedly over several hours, we find that $T_1$, $T_2^*$, and $T_{2,\mathrm{echo}}$ fluctuate in time, as shown in Fig.~\ref{fig:single-qubit}(d). The fluctuations are qualitatively similar to those commonly observed in angle-evaporated aluminum qubits \cite{van_harlingen_decoherence_2004,klimov_fluctuations_2018, burnett_decoherence_2019}. The time-averaged $T_1 = \SI{30}{\micro\second}$ corresponds to a qubit quality factor $Q = 2 \pi f T_1$ of one million.

Figure~\ref{fig:single-qubit}(e) shows the qubit level structure probed by microwave spectroscopy. At low drive powers, we observe a response at $f_{ge} = \SI{5.17}{\giga\hertz}$ corresponding to the transition from the ground to the excited state. When increasing the drive power, we see two further peaks appear, corresponding to the transition between the first and second excited states at $f_{ef} = \SI{5.114}{\giga\hertz}$ and the two-photon transition from the ground to the second excited state at $f_{gf}/2 = \SI{5.142}{\giga\hertz}$. This confirms the transmon-like level structure of our double-junction qubit, with an extracted anharmonicity of \SI{-56}{\mega\hertz}. The anharmonicity is approximately four times smaller than that of a conventional single-junction transmon with the same frequency and the same \SI{100}{\femto\farad} shunt capacitance. This is consistent with the intuition that the excitations remain transmon-like but now the phase drop is split over two junctions. Numerical simulations in Appendix~\ref{appendix:double-jj} confirm that the measured anharmonicity is consistent with $C_\mathrm{S} = 100\ \mathrm{fF}$ and junction capacitance on the order of 10 fF. Ten femtofarads in turn matches the capacitance expected for a junction with an effective area of $(600\ \mathrm{nm} - 90\ \mathrm{nm} )^2$, with specific capacitance of roughly \SI{40}{\femto\farad/\micro\meter^2} corresponding to a very low-transparency aluminum oxide tunnel barrier \cite{lichtenberger_fabrication_1989, mamin_merged-element_2021}.
In the appendix, we also numerically investigate another compromise between anharmonicity and charge dispersion and find that an anharmonicity of -102 MHz with charge dispersion of \SI{4}{\kilo\hertz} is possible, while keeping the qubit frequency similar.

\begin{figure}[tb]
\includegraphics{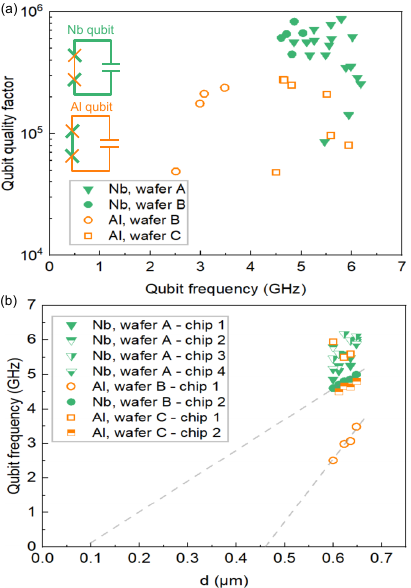}%
\caption{\label{fig:qubit-statistics} (a) Time-averaged qubit quality factors $Q$. Qubits with ground planes and the qubit island (filled markers) patterned into Nb2 produce on average $Q \sim 0.5$ million, while qubits with Al-BE (open markers) produce an average $Q$ of 0.2 million. Symbols indicate devices from different wafers.Inset shows two types of qubit variants with shunt capacitor and ground places patterned into Nb2 (top) or Al-BE (bottom). (b) Qubit frequency versus nominal junctions size $d$. Dashed gray lines are guides to the eye.}
\end{figure}

%\begin{figure}
%\includegraphics[width=\columnwidth]{Figure_Qubit freq_JJ_size.pdf}%
%\caption{\label{fig:qubit-freq-vs-d}Qubit frequency versus nominal junctions size $d$.}
%\end{figure}

Figure~\ref{fig:qubit-statistics}(a) shows time-averaged qubit quality factors of double-junction NAPA qubits from several chips on multiple wafers.  The wafers have different tunnel barrier oxidation parameters (see Appendix~\ref{appendix:devices-wafers}) and the nominal junction sizes range from 600 nm to 720 nm. Qubits with larger junctions are excluded because they have frequencies close to their readout resonators, and are thus limited by Purcell decay. On devices with qubit capacitors and ground planes patterned into Nb, we find relatively few outliers and an average quality factor of 0.5 million, with the best measured devices having a quality factor close to one million. Quality factors of Nb qubits show no correlation to junction size $d$ (see Fig.~\ref{fig:Q-vs-d} in Appendix~\ref{appendix:more-data}).

Al qubits, with capacitors patterned into Al and the small island between the two junctions made of Nb, on the other hand show a noticeably worse average quality factor of 0.2 million. 
Furthermore, there appears to be a trend in qubit lifetime decreasing rapidly as nominal junction size decreases below 650 nm (Fig.~\ref{fig:Q-vs-d} in Appendix~\ref{appendix:more-data}), although the range of junction sizes and the number of data points are too small for a definite conclusion.  
We have also measured single-junction qubits with equivalent nominal junction sizes from 600 nm to 1 $\mu$m, where we achieve quality factors of 0.3 million for the best device and an average Q of 0.2 million, again with no clear dependence on junction size (Appendix~\ref{appendix:more-data}). The single-junction qubits have a large parasitic series junction, with dimensions 6.4 $\mu$m $\times$ 3 $\mu$m, connecting the Josephson junction to the ground plane, which may limit qubit lifetimes \cite{lisenfeld_electric_2019}.

The lack of increased microwave loss with increased junction size would at first suggest that our qubit coherence is not limited by the tunnel junction itself or the spacer material, whose volume scales with the linear junction dimension. One would instead hypothesize that microwave losses in capacitors patterned into Al-BE are higher than in capacitors patterned into Nb2. 
However, we find similar low-power microwave loss in $\lambda/4$ coplanar waveguide resonators when comparing resonators patterned into Nb2 and Al-BE.
We measure such reference resonators on separate test chips from wafers A and C. In the low-photon-number limit, the internal quality factors are approximately 0.5 to 0.8 million regardless of layer, for resonators with 20 $\mu$m (10 $\mu$m) center pin (gap) width (Appendix~\ref{appendix:more-data}). 

Fig.~\ref{fig:qubit-statistics}(b) shows that qubit frequencies within a single chip scale approximately linearly with the nominal junction dimension, as expected, with the frequencies mostly falling within $\pm100\ \mathrm{MHz}$ of the linear trend. 
However, we find additional variation of roughly $\pm500\ \mathrm{MHz}$ from chip to chip, in particular in the data for wafer A that we have most data for.
We attribute this chip-to-chip scatter to nonuniformity of our deposition and etching processes on the full-wafer scale, as well as a strong local loading effect during etching.
The loading effect in reactive etching processes can lead to a large reduction in the etch rate when a large amount of material is etched in a given location. 
We see this loading effect most clearly in wafer B, where we fabricate both Nb qubits and Al qubits on the same wafer, with a checker pattern of 20~mm $\times$ 20~mm shots that alternate between Nb qubits and Al qubits. The challenge with this checker pattern is that almost all aluminum is etched away in the first patterning step from the Nb qubit shots, while almost no aluminum is etched from the Al qubit shots. This implies that the etch rate for the Al qubit shots is far higher than for the Nb qubit shots, due to the loading effect. We believe this leads to heavy overetching in the Al qubit shots and explains why wafer B Al qubit frequencies in Fig.~\ref{fig:qubit-statistics} are nearly a factor of two lower than Nb qubit frequencies, and the effective linewidth reduction indicated by the crossing of the dashed gray line is roughly 450 nm for those data points.

The loading effect could also lead to different amounts of etch residues in the vicinity of the qubit junctions in comparison to the reference resonators. This is one hypothesis for the difference in Nb and Al qubit performance despite similar reference resonator performance. However, this is speculative and we have not been able directly observe such residues. Despite lacking a satisfactory explanation for the dominant T1-limiting mechanism, if we assume pessimistically that all losses are related to the junctions, we can give an upper bound of order $10^{-5}$ for the effective loss tangent of the NAPA junctions, based on $1/Q \sim 10^{-6}$ and junction participation ratio $p_J \approx C_J/2C_\mathrm{S}$, which is approximately 5\% for the double-junction qubits.  The factor of two in the denominator is due to the junction capacitances being in series, and can also be seen in the definition of the effective shunt capacitance $C_\Sigma$ in Appendix~\ref{appendix:double-jj}.

In dc measurements shown in Appendix~\ref{appendix:dc}, we observe a subgap leakage conductance of roughly $1/2.5\ \mathrm{M}\Omega \approx 4 \times 10^{-3} \times R_\mathrm{N}^{-1}$ above critical temperature of Al, hinting at possible subgap states present in the Nb electrode of the tunnel junction. The appendix also shows that the subgap conductance does not necessarily improve when moving to millikelvin temperatures, although peaks of subgap current only allow to estimate an upper bound for the true leakage relevant to qubits from the dc data. Resonances of the uncontrolled electromagnetic environment of our test junctions are the most plausible explanation for such peaks~\cite{holst_effect_1994}. High subgap leakage is often observed in Nb-based tunnel junctions, while Al-based junctions can exhibit much lower values of leakage current \cite{saira_vanishing_2012}. 
In Ref. \cite{hatinen_efficient_2024}, dc measurements of much more transparent Al-Nb junctions, fabricated with the SWAPS process, exhibit a much lower \textit{ratio} of subgap leakage conductance to normal-state conductance. While the absolute leakage conductance demonstrated in Ref. \cite{hatinen_efficient_2024} is not lower, the lower ratio hints at the possibility of obtaining even lower leakage currents for NAPA junctions in the future.

% 2.5Mohm / ( 1/((2pi 5 GHz) * 100 fF) )  ~= 7850
At first sight, a parallel real conductance of $1/2.5\ \mathrm{M}\Omega$ should limit qubit quality factor to just $2.5\ \mathrm{M}\Omega/Z_{\mathrm{c},\mathrm{trans.}} \sim 10^4$, where $Z_{\mathrm{c},\mathrm{trans.}} \approx 320\ \Omega$ is the characteristic impedance of our transmon-like qubits. However, the relevant energy scale for qubit $T_1$ is $h(\SI{5}{\giga\hertz}) \approx \SI{20}{\micro e\volt}$, which is only one tenth of the superconducting gap of Al. If Al-BE has a good superconducting gap with no subgap states, there are no states for electrons in the possible subgap states of Nb2 to tunnel into, and they cannot exchange energy with the qubit degree of freedom. 
In future experiments, this relationship between qubit coherence and zero-bias conductance could be studied in more detail by carefully designing the microwave environment of the dc test structures to suppress resonances with the environment, while further theoretical work could provide quantitative predictions for the relation between subgap states, quasiparticles and qubit coherence \cite{Gustavsson2016, Bretheau2013, Zgirski2011}. 

\section{Conclusions}

In conclusion, we introduced a method for creating native-oxide-passivated (NAPA) Josephson junctions where the tunnel barrier is defined by an Al-Al$\mathrm{O}_x$-Nb trilayer and patterning is done subtractively, without at any point exposing the sidewall passivation to aggressive oxide removal steps. We demonstrated hallmarks of Josephson junction behavior in both dc characteristics and by showing that qubits utilizing NAPA junctions have transmon-like level structure.  For our best qubits, we find quality factors up to one million, slightly above the best reported for trilayer-type qubits, or any qubit utilizing Nb as a junction material. The limiting factor for qubit quality factor is not clear based on these proof-of-principle results, but it is likely that improved coherence times can be achieved in the future by optimizing the etch recipes, by creating the shunt capacitors in separate processing steps, or by reducing junction dimensions. To further elucidate the origin of decoherence, Ramsey noise spectroscopy \cite{Bylander2011} could be employed to disentangle low frequency noise from the high frequency dielectric loss inferred from $T_1$.
Furthermore, since the processing steps for NAPA junctions are compatible with modern fabrication tools and processes, we believe that frequency targeting can be dramatically improved in the future by following standard industry practices. 
Therefore, our results pave the way forward toward manufacturing superconducting qubits at scale. Finally, NAPA junctions may also offer an interesting platform for future studies of the impact of extreme gap asymmetry on quasiparticle tunneling.

%\subsection{}
%\subsubsection{}
\begin{acknowledgments} 
We acknowledge Tuomo Honkaharju and Jarmo Roine for technical help in the sample fabrication in VTT and OtaNano Micronova cleanroom facilities, and Harri Pohjonen and Isabel Gueissaz-Mattelmäki for technical assistance. We also thank Tero Heikkilä for valuable discussions.
The research was funded by the European Union’s Horizon RIA, EIC and ECSEL programmes under Grants  No. 824109 European Microkelvin Platform (EMP), No. 101113086 SoCool, No. 101007322 MatQu, No. 899558 aCryComm and No. 101113983 Qu-Pilot. We also acknowledge financial support of Research Council of Finland through projects No. 322580 ETHEC, No. 356542 SUPSI, the QTF Centre of Excellence project No. 336817, and Business Finland through Quantum Technologies Industrial (QuTI) No. 128291 and Technology Industries of Finland Centennial Foundation and Chips JU project Arctic No. 101139908.% TODO: ask Mika to double-check still
\end{acknowledgments}

\appendix
\section{Fabrication details}
\label{appendix:fabrication}
The fabrication proceeds as follows: First, we deposit the Al-Al$\mathrm{O}_x$-Nb trilayer by dc magnetron sputtering on a 150 mm high-resistivity silicon wafer, with a thickness of 100 nm for the bottom Al electrode and 100 nm for the Nb counter electrode. The aluminum oxide tunnel barrier is formed in situ by exposing the bottom aluminum electrode to 30 to 150 Torr of pure oxygen for 15 to 30 minutes, depending on the wafer. Relative to typical Manhattan-style qubit junctions, the oxygen pressure is high because of the large junction size and the larger superconducting gap of the niobium counter electrode. 

Next, we pattern the trilayer using optical lithography and fluorine-based reactive ion etching (RIE) to etch Nb-CE followed by chlorine-based RIE to etch Al-BE, timed such that overetch into silicon remains small ($\sim$20 nm). A native oxide forms on the etched sidewalls of both Al-BE and Nb-CE upon exposure to ambient conditions. After resist removal, we sputter Nb2a, without any oxide removal steps. We then spin photoresist and optically pattern a 200 nm $\times$ 500 nm mask opening on top of the junction and use fluorine-based RIE to create a via that goes through Nb2a and penetrates partially into Nb-CE.
Preceding sputtering of Nb2b,
we use in-situ Ar-based ion milling to remove the native oxides from Nb2a and the part of Nb-CE exposed through the via. 
Finally, we use optical lithography to pattern the top electrode and etch through Nb2b, Nb2a and Nb-CE using fluorine-based RIE, following the same approach as in conventional SWAPS junction fabrication \cite{gronberg_fabrication_2007}. 
This last (fluorine based) etch is selective to Al-BE due to the formation of non-volatile AlF$_3$, thus effectively stopping at Al-BE. However, it is not selective between niobium and silicon, and thus etches into the substrate by a significant amount ($\sim$120 nm) in areas not covered by Al-BE. All structures are patterned optically using an i-line stepper, and thus we limit the minimum nominal junction size to 500 nm $\times$ 500 nm. There is, however, no great barrier to going to smaller junction sizes in the future, with either more advanced optical lithography or electron beam lithography.
\section{\label{appendix:double-jj}Double-junction qubit}

\begin{figure}[htb]
\includegraphics{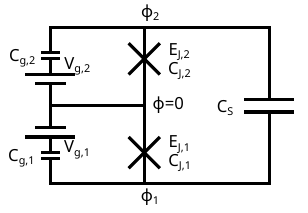}%
\caption{\label{fig:double-JJ-circuit}Electrical circuit representation of a double-junction qubit.}
\end{figure}

Figure~\ref{fig:double-JJ-circuit} shows an electrical circuit representation of the double-junction qubits studied here. Mathematically, it is convenient to define the island between the two junctions as having ($\phi=0$) and define $\phi_\pm = (\phi_2 \pm \phi_1)/2$, where $\phi_1$ ($\phi_2$) is the phase drop across the junction from the small island to the lower (upper) node. Under the simplifying assumption of identical junctions, one can then show that the Hamiltonian is

\begin{align}
H= & \frac{e^2}{2C_{\Sigma}}\left(n_{-}-n_{\text{g},-}\right)^{2}+\frac{e^2}{2C_{\Delta}}\left(n_{+}-n_{\text{g},+}\right)^{2} \nonumber \\
 & -2E_{\text{J},1}\cos(\frac{2\pi\phi_{+}}{\phi_{0}})\cos(\frac{2\pi\phi_{-}}{\phi_{0}}),
\end{align}
where
$\phi_{0}=2e/h$, 
$C_{\Delta}=C_{\text{J},1}+C_{\text{J},2}+C_{\text{g},1}+C_{\text{g},2}$, $C_{\Sigma}=4C_{\text{S}}+C_{\Delta}$, 
$n_{\pm}=n_{2} \pm n_{1}$ are Cooper pair number operators,
and $n_{\text{g},\pm}$ parametrize offset charges arising from $V_{\text{g},1}$ and $V_{\text{g},2}$.
We have chosen to use $C_\Sigma$ for the capacitance associated with the \textit{difference} operator, because we wish to highlight the similarity of $\phi_-$ excitations to those of a conventional transmon qubit, where the same symbol is commonly used to parametrize the effective shunt capacitance.

There are multiple ways to view this circuit: a transmon qubit with the junction replaced by a single Cooper pair transistor, two frequency-degenerate fixed-frequency transmons ultrastrongly coupled to each other by $C_\mathrm{S}$, a flux qubit in the limit of vanishing small-junction critical current but large shunt capacitance, or a quantronium circut in the limit of a vanishingly small inductive shunt element \cite{Vion2002}. Note that neither $\phi_+$ nor $\phi_-$ can be considered to be effectively classical for the parameters relevant here. Rather, the first excited states correspond to excitations along $\phi_-$ for the experimental parameters here, but already the fifth excited state corresponds to an excitation along $\phi_+$, as shown below.

Numerically, it is straightforward to simulate the circuit, directly in the basis of $\phi_1$ and $\phi_2$, using the flux qubit circuit model built into the scQubits Python package \cite{groszkowski_scqubits_2021,chitta_computer-aided_2022}, with the small junction Josephson energy set to a vanishingly small value. Below, we express capacitances in femtofarads and convert them in the simulations to charging energies by dividing $e^2/2$ by the capacitance in question.

\begin{figure}[tb]
\includegraphics{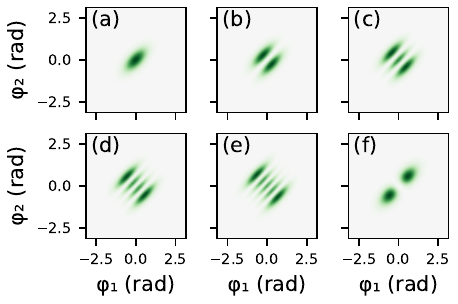}
\caption{\label{fig:double-JJ-wavefunctions}Wavefunctions of the six lowest eigenstates of the double-junction qubit circuit in Fig.~\ref{fig:double-JJ-circuit}, with $C_\mathrm{S} = 100\ \mathrm{fF}$, $C_{\mathrm{J},1} = C_{\mathrm{J},2} = 10\ \mathrm{fF}$, $E_{\mathrm{J},1} = E_{\mathrm{J},2} = 40\ \mathrm{GHz} \times h$ and zero gate charges.
Here, $\varphi_{i} = {2\pi\phi_{j}}/{\phi_{0}}$.
The intensity of the green color corresponds to the wavefunction squared, in arbitrary units. The transition frequencies, from the ground state, are (b) 5.165 GHz, (c) 10.276 GHz, (d) 15.331 GHz, (e) 20.331 GHz, (f) 23.436 GHz.}
\end{figure}

Figure~\ref{fig:double-JJ-wavefunctions} shows the wavefunctions of the first eigenstates and lists their energies, for the experimentally relevant parameters. The simulated anharmonicity is \SI{-54}{\mega\hertz} for a qubit frequency of 5.165 GHz, in line with the measured anharmonicity of \SI{-56}{\mega\hertz} for the 5.17 GHz qubit discussed in the main text.

%Figure~\ref{fig:double-JJ-fqubit-vs-ng}(b) shows
The numerical simulations show that the double-junction qubits studied here are extremely charge insensitive, with the first transition frequency having a variation of only 2 Hz over $n_{\mathrm{g},\pm}$. This extreme charge insensitivity can be understood by observing that the lowest energy states in Fig.~\ref{fig:double-JJ-wavefunctions} correspond to the transmon-like $\phi_-$ excitations and comparing the Hamiltonian to that of a standard transmon \cite{koch_charge-insensitive_2007} with Josephson energy $E_\mathrm{J}$ and charging energy $E_\mathrm{C}$. Under the heavily-simplified approximation $\phi_+ \approx 0$, the Hamiltonian here looks like that of a conventional transmon but with $2E_{\mathrm{J},1}$ in place of $E_\mathrm{J}$ and approximately $(e^2/2C_\mathrm{S})/4$ in place of $E_\mathrm{C}$. Note the factor of four in the effective charging energy, arising from the additional factor of four in the definition of $C_\mathrm{\Sigma} \approx 4C_\mathrm{S}$ for the double-junction qubit. In order to maintain comparable qubit frequency for the same $C_\mathrm{S}$, $2E_{\mathrm{J},1}$ must therefore be chosen as $4E_\mathrm{J}$. The simulated qubit frequency of 5.15 GHz is indeed relatively close to $\sqrt{8 (2E_{\mathrm{J},1}) (e^2/2C_\mathrm{S})/4}\approx5.57\ \mathrm{GHz}$, as you would expect for a transmon. The low charging energy explains the low anharmonicity, but also note that the ratio of the effective Josephson energy $2E_{\mathrm{J},1}=4E_\mathrm{J}$ to charging energy $(e^2/2C_\mathrm{S})/4 = E_\mathrm{C}/4$ is sixteen times larger than $E_\mathrm{J}/E_\mathrm{C}$ of the conventional transmon, with the same $C_\mathrm{S}$ and comparable frequency. This explains the extremely low value of simulated charge sensitivity.

Unfortunately, approximating $\phi_+$ as zero is too simplistic and in reality sensitivity to $n_{\mathrm{g},+}$ dominates charge dispersion. Nevertheless, we can partially recover anharmonicity by sacrificing some charge insensitivity. In particular, if we decrease $C_\mathrm{S}$, $E_{\mathrm{J},1}$ and $E_{\mathrm{J},2}$ by 45\%, we achieve an anharmonicity of -102 MHz and a charge dispersion of \SI{4}{\kilo\hertz}, with the qubit frequency changing only little to 4.94 GHz. For these parameters, it is the fourth, rather than the fifth, excited state at 16.957 GHz that corresponds to the first $\phi_+$ excitation.

\section{\label{appendix:more-data}Additional data}

\begin{figure}[tb]
\includegraphics{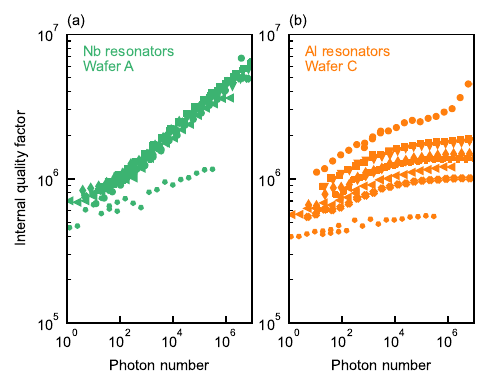}%
\caption{\label{fig:resonators}Resonator internal quality factors as functions of circulating photon number for (a) resonators patterned into Nb2 on wafer A and (b) into Al-BE on wafer C. Markers indicate resonators with different frequencies on the same chip. }
\end{figure}

Figure~\ref{fig:resonators} shows the measured internal quality factors of $\lambda/4$ coplanar waveguide resonators patterned in Al-BE or Nb2, measured on separate test chips on wafers A and C (wafer details are in Appendix~\ref{appendix:devices-wafers}). The test chips have eight resonators with frequencies between 4.5 GHz and 7.7 GHz coupled to a common feedline. The CPW center trace has a width of 20 $\mu$m and the gap between the center trace and the ground plane is 10 $\mu$m. For the Nb resonators, both the low-power internal quality factors and their photon-number dependence are typical of Nb resonators, as you can see by comparing Fig.~\ref{fig:resonators}(a) to, e.g., the planar Nb reference shown in Fig.~2(a) of Ref.~\cite{grigoras_qubit-compatible_2022}, which also shows internal quality factor of four million at $10^6$ photons, for the same CPW gap and width. It is perhaps even surprising that the resonators patterned into Nb2 perform so well, even though both the substrate-metal and the substrate-air interfaces have seen significant additional processing due to deposition and removal of a trilayer, before Nb2 deposition and its patterning. The Al resonators show similar low power quality factors but much weaker power dependence and considerable scatter in the power-independent high-power quality-factor. It is typical for Al resonators for the high-power quality factor to saturate to lower values than for Nb resonators (e.g. 3 million in  Ref.~\cite{biznarova2024mitigation}, but the scatter suggests that there is room to improve and is, qualitatively, in line with the hypothesis of non-uniform etch residues. Nevertheless, the high low-power internal quality factors ($\ge$0.4 million) do not straightforwardly explain the observed T1 times.

\begin{figure}[tb]
\includegraphics{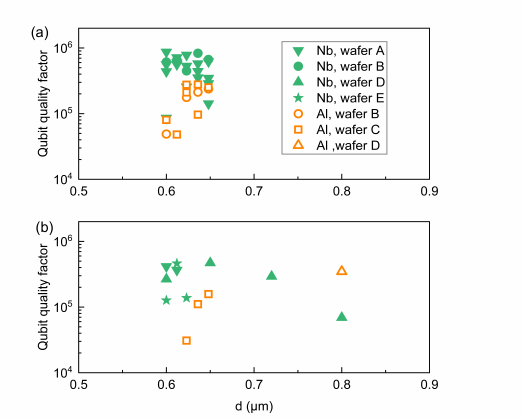}
\caption{\label{fig:Q-vs-d}Qubit quality factors of (a) double-junction qubits and (b) single-junction transmons, as functions of nominal Josephson junction size $d$. Green (orange) markers indicate devices with qubit electrodes patterned into Nb2 (Al-BE). Different symbols correspond to different wafers. }
\end{figure}

In Figure~\ref{fig:Q-vs-d}, we plot measured quality factors versus nominal Josephson junction size $d$. The data points in panel (a) are the same as in Fig.~\ref{fig:qubit-statistics}, except for the horizontal axis. The qubit frequencies for the single-junction transmons in panel (b) range from 4 GHz to 9 GHz, which means that some qubits have their frequency above the corresponding readout resonator. The devices with nominal dimension above 0.9 $\mu$m are flux-tunable transmons with two junctions in parallel forming a SQUID loop, and the given dimension is the size of an effective single junction with the same nominal area. The flux-tunable transmons were characterized at nominally zero flux threading the SQUID loop. We find average qubit $Q$ of 0.3 million for single-junction Nb qubits and 0.2 million for single-junction Al qubits, with no clear dependence on $d$ in case of Nb qubits. Al qubits on the other hand show a trend of decreasing quality factor with decreasing $d$ below 650 nm, more clearly so in case of wafer B and less clearly in case of wafer C.

\begin{figure}[tb]
\includegraphics{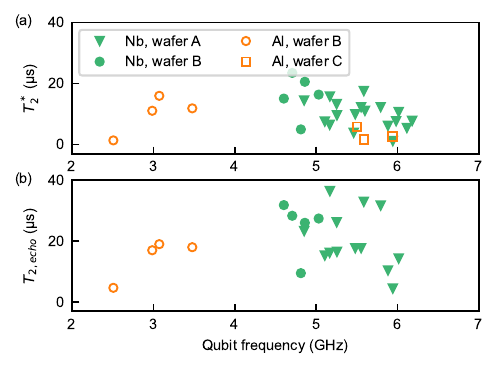}
\caption{\label{fig:T2}(a) Ramsey decoherence time $T_2^*$ and (b) Hahn echo decoherence times $T_{2,\mathrm{echo}}$ for the devices shown in Fig.~\ref{fig:qubit-statistics} of the main text. Filled markers indicate devices with qubit electrodes in the Nb2 layer and open markers in Al-BE. Symbols indicate devices from different wafers. }
\end{figure}

In Figure~\ref{fig:T2}, we show the time-averaged $T_2$ times of the same devices for which qubit quality factors are shown in Fig.~\ref{fig:qubit-statistics} in the main text. We find $T_2^*$ up to 25 $\mu$s and $T_{2,\mathrm{echo}}$ up to nearly 40 $\mu$s, with $T_{2,\mathrm{echo}}$ typically higher than $T_2^*$ by up to a factor of two. Note that the $T_2$ times were not characterized for all devices. The somewhat lower $T_2$ times observed in Al qubits are most likely explained by their lower $T_1$ compared to Nb qubits, although we believe that the $T_2$ times in general may be limited by the measurement setup.

\section{\label{appendix:measurement-setup}Measurement setup}

We measure qubits in a dilution refrigerator setup similar to that used in Ref. \cite{grigoras_qubit-compatible_2022}. The input line, used for both qubit drive and readout pulses, is attenuated by nominally 61 dB distributed among the different temperature stages, and filtered with low-pass filters and infrared filters at the base temperature stage. The qubit chips are wire-bonded to gold-plated copper sample holders, mounted inside magnetic shields. We use a a pair of microwave switches at the base temperature stage to allow for characterizing up to five chips in a single cooldown. There are minor changes in the filtering configuration between cooldowns, but we have not observed significant changes in qubit performance. The output signal is amplified by a traveling-wave parametric amplifier at base temperature and a high-electron-mobility transistor at 4 K, followed by further amplification at room temperature. At room temperature, we use a Zurich Instruments SHFQC for generating the readout and drive pulses and digitizing the readout signals.

\begin{table*}[ht]%[H] add [H] placement to break table across pages
\caption{\label{table:anharmonicity}Measured qubit frequencies, anharmonicities and extracted junction capacitances for single-junction transmon qubits with different junction sizes but similar Xmon geometry. The $T_1$ coherence data is time-averaged over 4-8 hours.}
\begin{ruledtabular}
\begin{tabular}{p{2.5cm}p{2cm}p{2.5cm}p{2cm}p{2cm}p{2cm}p{2.5cm}}
Nominal junction size ($\mu$m) & Wafer & Qubit frequency (GHz) & $T_1$ \SI{}{(\micro\second)}& Anharmonicity (MHz) & $E_C/h$ (MHz) & Junction capacitance (fF) \\ \hline
0.8 & D & 4.389 & 12.6 & -176  & 162 & 22 \\
0.612 & E & 4.848 & 15.2 & -208 & 185 & 7 \\
0.2 & Al-Al$\mathrm{O}_x$-Al reference & 4.294 & 140 & -220 & 196 & 1
\end{tabular}
\end{ruledtabular}
\end{table*}

\section{\label{appendix:devices-wafers}Devices and wafers}

We have characterized devices from five different wafers, A, B, C, D, and E. The main difference between the processing is the tunnel barrier oxidation parameters: Most wafers were oxidized for 15 minutes at 30 Torr, while wafer D was oxidized for 30 minutes at 150 Torr. The SEM and TEM images shown in Figures \ref{fig:schematic} and \ref{fig:junction} are taken from wafer A. The room temperature junction resistance data in Fig.~\ref{fig:junction} is from wafer A, as is the qubit device characterized in Figure~\ref{fig:single-qubit}. The low-temperature dc measurements shown in Fig.~\ref{fig:junction} and Appendix~\ref{appendix:dc} are from wafer B. Wafers A and E have only Nb qubits, wafer C has only Al qubits, and wafers B and D have both in the checker pattern mentioned in the main text.

All characterized qubits have a simulated shunt capacitance of approximately \SI{100}{\femto\farad}. The measured qubit-readout resonator couplings $g/2\pi$ are between 50 MHz and 90 MHz, while the coupling between the readout resonator and the feedline $\kappa/2\pi$ varier from 100 kHz to 300 kHz.
Most measured qubits use capacitors in the Xmon geometry with 20 $\mu$m arm width and 20 $\mu$m gap between the qubit island and the ground plane, although a few devices have arm widths and gaps ranging from \SI{10}{\micro\meter} to \SI{80}{\micro\meter}. Our best-performing device, shown in Fig.~\ref{fig:single-qubit} of the main text, has a \SI{10}{\micro\meter} arm width and gap. We have not observed a consistent dependence between the Xmon geometry and qubit performance. The summary plots shown in Fig.~\ref{fig:qubit-statistics} of the main text as well as Appendix~\ref{appendix:more-data} include all characterized fixed-frequency qubits. 

To estimate the junction capacitance, we measure anharmonicities from single-junction qubits with different nominal junction sizes, shown in Table \ref{table:anharmonicity}. For reference, we also include a device with angle-evaporated Al-Al$\mathrm{O}_x$-Al junctions and Nb electrodes, with the same Xmon geometry. We then find the charging energies $E_C$ matching measured anharmonicities and qubit frequencies using the scQubits package, calculate the corresponding total capacitance, and subtract the simulated external shunt capacitance to obtain the capacitance contributed by the junction. The contribution of the junction capacitance in the Al-Al$\mathrm{O}_x$-Al reference qubit is near-negligible. We find junction capacitances of \SI{7}{\femto\farad} for a nominally 612 nm wide junction (22 fF for 800 nm wide junctions). These values are consistent with an expected capacitance per area of roughly \SI{40}{\femto\farad/\micro\meter^2}, considering that devices from different wafers likely have different true junction areas for a given nominal size due to variations in lithography and etching. For the simulations of double-junction qubits described in Appendix~\ref{appendix:double-jj}, we use 10 fF as an estimate for the typical capacitance per junction in measured devices.

\section{\label{appendix:dc}Low-temperature dc measurements}

We performed low-temperature dc measurements on separate test structures from wafer B. The devices are mounted at the mixing chamber stage of a dry dilution refrigerator equipped with radiation shielding and filtering \cite{kemppinen_long_2011, saira_vanishing_2012}. We apply a voltage bias at room temperature and measure both the current and voltage across the device in a four-probe configuration with room-temperature transimpedance and voltage amplifiers. The measurement lines have a resistance of approximately 2 k$\Omega$ per line. 

\begin{figure}[b]
\includegraphics{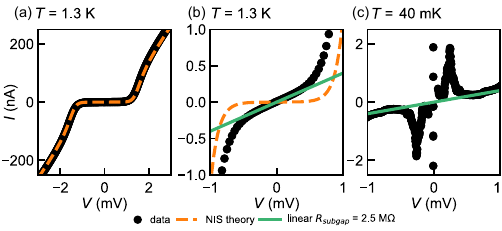}
\caption{\label{fig:dc-subgap}
(a) Current-voltage characteristics of a test junction at 1.3 K (markers), as well theory prediction for a normal--insulator--superconductor junction (dashed orange line). % Theory curve is just the basic NIS formula at 1.3 K with the given Delta and R (and a dynes gamma that's negligible on this scale for convenience (read: Elsa didn't want to type up more than one formula))
(b) Zoom of the data shown in panel (a), along with a linear slope with $R_\mathrm{subgap}=2.5$ M$\Omega$ (solid green line).
(c) Data measured at 40 mK (markers) with the same voltage range as in panel (b), with the same linear slope (solid green line) as a guide to the eye.
}
\end{figure}

In Figure~\ref{fig:dc-subgap}(a), we show measured current-voltage characteristics of a junction with nominal $d = 629$ nm at 1.3 K, where the Al bottom electrode is in the normal state. We find approximately zero current up to a voltage of approximately 1.4 mV, as expected for tunneling in a normal--insulator--superconductor (NIS) junction \cite{giazotto_opportunities_2006}. On this scale, we find good agreement with simulations with normal-state resistance $R_\mathrm{N} = $10 k$\Omega$, measured at 10 K, and niobium superconducting gap $\Delta_{\mathrm{Nb}} = 1.42$ mV, extracted as shown below. On a smaller scale [Fig.~\ref{fig:dc-subgap}(b)], however, we find a linear-in-$V$ subgap slope corresponding to a resistance $R_\mathrm{subgap} = 2.5$~M$\Omega$, with a far larger conductance than expected from thermal excitations alone. The subgap current in a NIS junction is often characterized with the phenomenological Dynes parameter $\gamma$ \cite{dynes_direct_1978}, and here we can estimate $\gamma = R_\mathrm{N} / R_\mathrm{subgap} = 4 \times 10^{-3}$. 

\begin{figure}[!bt]
\includegraphics{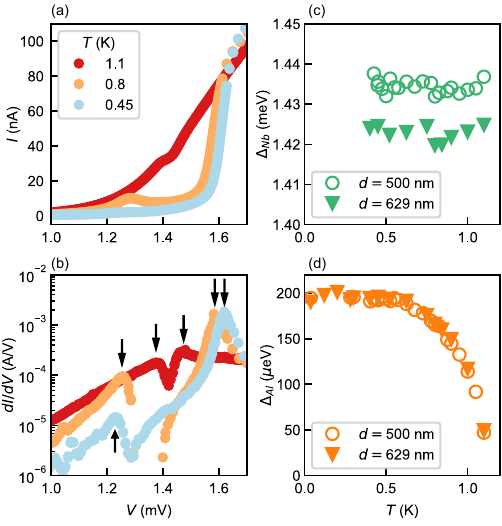}%
\caption{\label{fig:dc-gap-extraction}
(a) Measured current-voltage characteristics and  for the same test junction as in Fig.~\ref{fig:dc-subgap} at several temperatures. (b) Numerical differential conductance calculated from the data in panel (a), with arrows indicating peaks occurring at $|V| = \Delta_{\mathrm{Nb}} \pm \Delta_{\mathrm{Al}}$. 
(c) Extracted superconducting gaps of niobium $\Delta_{\mathrm{Nb}}$ and (d) aluminum versus the bath temperature.
}
\end{figure}

At 40 mK [Fig.~\ref{fig:dc-subgap}(c)], we observe a supercurrent branch, as shown in Fig.~\ref{fig:junction}(f) of the main text, features corresponding to the sum and difference of the superconducting gaps of Nb and Al (shown in Fig.~\ref{fig:dc-gap-extraction}), as well as additional subgap structure. Outside the supercurrent branch, the IV is not hysteretic. The features at $|V|<0.5$ mV are likely due to resonances in the microwave environment of the junction~\cite{holst_effect_1994}, impeding characterization of the subgap in the region $|V| < \Delta_{\mathrm{Al}}/e = 0.2 $~mV relevant for qubit operation. We observe qualitatively similar behavior in another device with $d=629$ nm and $R_\mathrm{N} = 20$ k$\Omega$, finding $R_\mathrm{subgap} = 3.5$~M$\Omega$ at 1.3 K, corresponding to $\gamma = 6 \times 10^{-3}$.

%% This sentence is probably not really needed in the end, but it's a piece of data I have
%We note that the resistance of our Al-Nb junctions increases by 20 to 50 percent at 10 K compared to room temperature, which is explained by suppression of thermally activated transport above the tunnel barrier and the freezing out of parallel conductance through the high-resistivity silicon substrate. 

\clearpage
To extract the superconducting gaps of Nb and Al, $\Delta_{\mathrm{Nb}}$ and $\Delta_{\mathrm{Al}}$, respectively, we numerically calculate the differential conductance $dI/dV$ from the measured current $I$ and voltage $V$ and extract the voltage at which peaks occur in the differential conductance, as shown in Fig.~\ref{fig:dc-gap-extraction}(a--b). We identify the peaks occurring at $|V| > 1.4 $~mV as corresponding to $(\Delta_{\mathrm{Nb}} + \Delta_{\mathrm{Al}})/e$ and those at $1.0$~mV$ < |V| < 1.4$~mV as corresponding to $(\Delta_{\mathrm{Nb}} - \Delta_{\mathrm{Al}})/e$. The peaks at $(\Delta_{\mathrm{Nb}} - \Delta_{\mathrm{Al}})/e$ arise from aligning the energy levels of thermally excited quasiparticles in Al with the gap edge of Nb~\cite{muhonen_micrometre-scale_2012}. We average the peak positions found at positive and negative voltages for a given bath temperature and extract the gaps versus bath temperature, as shown in Fig.~\ref{fig:dc-gap-extraction}(c--d). 
Below 0.4 K, the peaks at $\Delta_{\mathrm{Nb}}-\Delta_{\mathrm{Al}}$ are no longer visible due to the small number of quasiparticles in Al. However, we can still extract the peaks at $\Delta_{\mathrm{Nb}} + \Delta_{\mathrm{Al}}$ and extract $\Delta_{\mathrm{Al}}$ by assuming that $\Delta_{\mathrm{Nb}}$ stays constant at low temperatures. % We can also leave this sentence out and just cut the plots at 0.4 K
We find $\Delta_{\mathrm{Nb}} \approx 1.4$ mV for both devices. For aluminum, we find $\Delta_{\mathrm{Al}} = 200~\mu$eV at the lowest measured temperatures, with little variation below 0.5 K.

% tables should appear as floats within the text
% The ruledtabular enviroment adds doubled rules to table and sets a
% reasonable default table settings.
% Use the table* environment to get a full-width table in two-column
% Add \usepackage{longtable} and the longtable (or longtable*}
% environment for nicely formatted long tables. Or use the the [H]
% placement option to break a long table (with less control than 
% \begin{table}%[H] add [H] placement to break table across pages
% \begin{ruledtabular}
% Lines of table here ending with \\
% \end{tabular}
% \end{ruledtabular}
% \end{table}

% Surround table environment with turnpage environment for landscape
% table
% \begin{table}
% \begin{ruledtabular}
% \end{tabular}
% \end{ruledtabular}
% \end{table}

%\bibliography{QHW_common_20250220.bib}
\bibliographystyle{apsrev4-2}
\bibliography{trilayer_qubits}
\end{document}